\documentclass[preprint,prc,aps,showpacs,showkeys,groupedaddress,floatfix]{revtex4-2}
\usepackage{epsfig}
\usepackage{dcolumn}
\usepackage{bm}
\usepackage{graphics}
\usepackage{graphicx}

\begin{document}

\title{Sensitivity analysis for the anomalous $tq\gamma$ couplings via $ \gamma q {\rightarrow} t {\gamma}$ subprocess in photon-proton collisions at the FCC-${\mu}$p}

\author{E. Alici}
\email[]{edaalici@beun.edu.tr} \affiliation{Department of Physics, Zonguldak Bulent Ecevit University, Turkey}

\begin{abstract}
In this study, we investigate anomalous flavour-changing neutral current (FCNC) interactions related to the top quark, particularly the $t\rightarrow q\gamma$ transition, within the Standard Model Effective Field Theory (SMEFT) framework. These rare processes are largely suppressed in the Standard Model (SM) and are strong indicators for new physics scenarios beyond the SM. In our study, we have analyzed the cross sections of the $ \gamma q {\rightarrow} t {\gamma}$ subprocess for two different center-of-mass energies, 17.3 TeV and 24.5 TeV, at the Future Circular Collider (FCC-${\mu}$p) collider. Our results derived from simulations show that the branching ratios for these processes can be reduced to the order of $10^{-5} $. These values provide significantly tighter limits than the current experimental limits. In light of these findings, it can be stated that the FCC-${\mu}$p collider offers significant potential for discovering new physics which represents a sensitivity improvement over the current experimental limits from CMS, enhancing sensitivity by approximately 30\%.Moreover, it has been demonstrated that higher centre-of-mass energies and integrated luminosity values significantly enhance the discovery potential of these rare processes. Accordingly, the results indicate that-energy colliders such as FCC-${\mu}$p are of critical importance in the search for new physics and in the study of the role of the top quark in the context of flavour physics. As a result, by providing a comprehensive theoretical analysis, this study contributes to the importance of FCNC processes in the search for new physics beyond the Standard Model and will shed light on future experimental and phenomenological studies.

Keywords: Top quark physics,Flavour-changing neutral current, photon-proton collisions, FCC muon-proton collider

\end{abstract}

\pacs{14.65.Ha}

\maketitle

\section{INTRODUCTION}\label{sec3}
The study of flavour-changing neutral currents (FCNCs) involving the top quark represents a significant area of research in particle physics, particularly in view of the potential they offer for exploring new physics phenomena beyond the Standard Model (SM). In the SM, the suppression of FCNC processes, which enable transitions between different flavours of quarks without changing their electric charge, is attributed to the Glashow-Iliopoulos-Maiani (GIM) mechanism \cite{1}. This suppression leads to the expected branching ratios for the FCNC decays of the top quark to be on the order of $10^{-16}$ to $10^{-14}$\cite{2,3}. On the other hand, various extensions to the SM may allow us to predict even advanced FCNC interactions, particularly those involving the top quark, the heaviest known elementary particle. The top quark has a mass close to the electroweak symmetry breaking scale, making its interactions specially more sensitive to potential new physics scenarios\cite{1}. Also, the rarity of FCNC processes in the SM means that they serve as a clean probe for new interactions that may be introduced in various Beyond-Standard Model (BSM) theories (which include various extensions of the SM). The theoretical framework surrounding FCNC processes is rich and varied, with numerous BSMs proposing different mechanisms that could enhance these transitions. At this point, it is of paramount importance to establish a clear understanding of the relationship between theoretical predictions and experimental findings in order to assess the viability of these models and their implications for the fundamental structure of matter. A particularly important mechanism in this context is the Standard Model Effective Field Theory (SMEFT). In light of these considerations, the researchs on exploration of flavour-changing neutral currents (FCNC) involving the top quark within the framework of the SMEFT has emerged as a topic of considerable attention in recent years.
The SMEFT provides a systematic methodology for analysing the effects of new physics by extending the SM with higher-dimensional operators, in particular those of dimension six, which can modify the interactions of the top quark and other fermions. This approach allows a more comprehensive examination of FCNC processes, which are in contrast highly suppressed in the SM due to the GIM mechanism. Here, as it is known with regard to FCNC interactions, the top quark serves as a crucial probe for new physics, due to its large mass and the associated Yukawa couplings.  In that respect, the SMEFT framework predicts that the branching ratios for FCNC decays such as  $t q\gamma$ can be significantly enhanced in various beyond-the-SM scenarios \cite{4,5,6,7,8,9,10,11,12}. Accordingly, phenomological and experimental studies have shown that the branching ratios for these processes could reach values on the order of $10^{-5} $ to $( 10^{-6} )$ under specific conditions\cite{13,14,15,16,17,18,19,20,21,22,23,cms,atlas}. 

Experimental searches for FCNC processes involving the top quark have been conducted, herein, at the Large Hadron Collider (LHC), with various analyses focusing on single top quark production in association with photons. In a recent study,, the CMS collaboration has reported limits on the branching ratios for FCNC decays  that experimental upper limits on the branching ratios (BR) have been achieved: $BR(t\rightarrow u\gamma)<0.95\times10^{-5}$ and $BR(t\rightarrow c\gamma)<1.51\times10^{-5}$ \cite {cms}. The sensitivity of these searches is anticipated to improve with the accumulation of more data and the implementation of advanced analysis techniques, such as those based on the SMEFT framework. Also, these searches are sensitive to the $t q\gamma$ coupling and can provide limits, which are crucial for constraining theoretical models that predict enhanced FCNC interactions. On the other hand, theoretical studies conducted within the SMEFT framework have also explored the implications of FCNC interactions for top quark production processes. For example, in one of them, the production of single top quarks via FCNC couplings was analysed, and it was revealed that cross sections for such processes can be developed in certain new physics models. These findings underscore the significance of FCNC processes as a means of probing the flavour structure of the underlying theory and searching for potential signals of new interactions. Moreover, the interplay between FCNC processes and other flavor-changing interactions has been a topic of considerable interest. The global fits within the SMEFT framework have been employed to extract constraints on the flavor-changing couplings of the top quark, thereby providing insights into the flavor structure of the theory and the potential connections between disparate flavor-changing processes. The combination of experimental data from top quark production and decay processes, in conjunction with theoretical predictions from the SMEFT, offers a powerful approach to understanding the dynamics of flavor in particle physics.

Considering the aforementioned significant literature, it can be said that the study of FCNC processes involving the top quark represents a crucial element of contemporary particle physics research. The suppression of these processes within the SM, coupled with the potential for enhancement in BSM scenarios, renders them an appealing target for experimental investigation. Ongoing and future experiments at high-energy colliders will continue to refine our understanding of top quark interactions and their implications for the broader framework of particle physics.

\section{FCC-$\mu$p collider}\label{sec3}

The study of FCNC transitions in top quark interactions is of paramount importance for gaining insight into the fundamental mechanisms of flavour physics and the potential existence of new particles or interactions. For instance, the coupling of the top quark to a photon tq$\gamma$ is of particular interest, as it can be probed through various production processes at high-energy colliders. 
At this point, future colliders, such as the Future Circular Collider (FCC), are expected to provide even greater sensitivity to FCNC processes. These colliders will also enable precision measurements of top quark properties and could potentially reveal indications of new physics through enhanced FCNC interactions. The pristine experimental environment and elevated luminosity of these colliders render them optimal for probing infrequent processes that are otherwise challenging to observe.
In this regard, the FCC project at CERN is positioned to become a pioneering endeavour in the field of high-energy physics, with the construction of a muon-proton collider representing a pivotal aspect of its planned developments\cite{fcc1}. It is well established that this collider aims to explore new physics beyond the SM by leveraging the unique properties of muons and protons in collisions. In this regard, it is expected that the FCC muon-proton collider will offer substantial advantages in terms of energy reach and luminosity, thereby becoming a crucial instrument for probing fundamental interactions and potential new particles.

One of the principal motivations for developing a muon-proton collider is the inherent superiority of muons as collision particles\cite{fcc2,fcc3,fcc4}. In contrast to electrons, muons possess a greater mass, enabling them to transfer more energy in collisions without substantial energy loss due to synchrotron radiation. This characteristic enables muon colliders to achieve higher centre-of-mass energies in a more compact configuration than electron colliders. Hence, the FCC muon-proton in the context of particle physics collider is expected to offer unique opportunities to discover new particles and study rare processes. Also, the collider's ability to produce high-energy collisions will facilitate the exploration of leptoquarks, heavy neutrinos, and other beyond-the-SM phenomena. The interaction of muons with protons also presents new avenues for flavor physics research, which could potentially elucidate the underlying mechanisms of flavor-changing processes.
The FCC muon-proton collider is part of a broader FCC initiative that includes various collider options, such as electron-positron and proton-proton colliders. This multifaceted approach permits a comprehensive exploration of particle physics, with each collider type contributing to different elements of the research programme. Moreover, the FCC signifies a substantial advancement in high-energy physics, notably in the investigation of photon interactions through the Weizsäcker-Williams (WW) approximation. The fundamental principle underlying these approximations is that a relativistic charged particle, such as an electron, muon or proton, generates an electromagnetic field that can be approximated as a stream of photons. These photons are considered to be "equivalent" due to the fact that they carry the same energy and momentum as the actual photons emitted by the charged particle. The WW method provides a means of calculating the flux of these equivalent photons, which can then be employed to predict cross-sections for a range of interactions, such as photon-photon and photon-proton collisions \cite{24,25,26,27}. These photons play a crucial role in various collision processes, including lepton-proton and photon-photon interactions, which are essential for probing the fundamental structure of matter and exploring new physics beyond the SM. In summary, it is evident that the FCC muon-proton collider with photon-proton modes has the potential to explore new frontiers in high-energy particle interactions and represents a significant advance in collider physics.

Consequently, the study of FCNCs involving the top quark is a vibrant field that bridges experimental and the potential for discovering new physics through enhanced FCNC interactions renders this area of research particularly compelling. As experimental techniques and theoretical models continue to advance, it is inevitable that our understanding of top quark FCNCs will deepen. This will provide insights into the fundamental nature of particle interactions and contribute to the quest for a more complete theory of particle physics.

This study examines the $tq\gamma$ transitions of anomalous FCNC interactions with single top quark production at FCC-$\mu$p for two different values of the centre-of-mass energy (17.3 and 24.5 TeV). At these centre-of-mass energies, the energy of the muon beam is 1500 GeV and 3000 GeV, while the energy of the proton beam is 50000 GeV.  In the process under consideration, the subprocess is the decay of a $ \gamma q {\rightarrow} t  \gamma $ , which subsequently decays into a  W boson and a b quark. Moreover, for the case where the W boson decays hadronically, in the final case our process becomes ${\mu} {p} {\rightarrow} {\mu} {\gamma} {p}  {\rightarrow} {\mu} t {\gamma}{\rightarrow} {\mu} W b {\gamma}{\rightarrow} {\mu}{} j j b {\gamma} $.

\section{Theoritical Framework}\label{sec3}

In the SMEFT, the Lagrangian is extended to include six-dimensional operators that respect the symmetries of the SM. These operators can be classified into a number of categories, including those that contribute to FCNC processes. For the top quark, relevant operators include those that couple the top quark to other quarks (up-type quarks) and photons. The effective Lagrangian can be expressed as follows\cite{28,29}:
\begin{eqnarray}
{\mathcal{L}}={\mathcal{L}}_{\textit{SM}}+\sum\frac{C_{i}O_{i}}{{\Lambda}^2}
\end{eqnarray}

In this context, ${\mathcal{L}}_{\textit{SM}}$ sets forth the SM Lagrangian, which encompasses four-dimensional operators. Meanwhile, $O_{i}$ elucidate six-dimensional operators pertinent to extended theories. Additionally, the symbol ${\Lambda}$ is used to denote the energy scale pertinent to extended theories, whereas the symbols $C_{i}$ are used to identify the Wilson couplings.
In this paper, we opt to employ the terms of the new physics, which encompasses the six-dimensional operators that describe the FCNC couplings between the top quark, q(u quark and c quark), and the photon.\cite{29},

\begin{eqnarray}
O^{ij}_{uW}=(\bar{q}_{Li}{\sigma^{\mu\nu} }{\tau}^{I}{u}_{Rj})\tilde{\phi}{}W^{I}_{{\mu}{\nu}}
\end{eqnarray}
\begin{eqnarray}
O^{ij}_{uB\phi}=(\bar{q}_{Li}{\sigma^{\mu\nu} }{u}_{Rj})\tilde{\phi}{}B_{{\mu}{\nu}}
\end{eqnarray}

In this formulation, the indices i and j are used to represent the flavour of the quark. The left-handed quark doublet is represented by the symbol ${q}_{Li}$, while the right-handed quark singlet is represented by ${u}_{RJ}$. The Pauli matrices are defined by ${\tau}^{I}$. Additionally, $\tilde{\phi}$ symbol is used to denote a complex conjugate $i{\tau}^{2}\phi^{*}$, where $\phi$ describes the SM Higgs doublet. Here, it is a readily straightforward process to extend the SM Lagrangian with the FCNC couplings at the top of tq${\gamma}$. The resulting Lagrangian is given by\cite{28,29}:
\begin{eqnarray}
{\mathcal{L}_{FCNC}}=\frac{g_{e}}{2m_{t}}\sum_{q=u,c}\bar{q}{\sigma_{\mu\nu} }({\lambda}_{qt}^{R}P_{R}+{\lambda}_{qt}^{L}P_{L})t A^{{\mu}{\nu}}+h.c
\end{eqnarray}

In the equation mentioned above, $\sigma_{\mu\nu}$ is defined as $[\gamma_{\mu},\gamma_{\nu}]/2  $, where $\gamma_{\mu}$ and $\gamma_{\nu}$  represent the Dirac matrices. The symbol $g_{e}$ 
denotes the electromagnetic coupling constant. Additionally, $P_{R}$ and $P_{L}$ refer to the right-handed and left-handed projection operators, respectively. The parameters $ {\lambda}_{qt}^{R}$ and $ {\lambda}_{qt}^{L}$ represent the dimensionless real constants that include the Wilson coefficients of the effective operators, describing the anomalous coupling between a top quark, a photon, and an up-type quark (u or c).
\begin{eqnarray}
\lambda^{L}_{qt}=\frac{\sqrt{2}}{e}[s_{w}C^{3j*}_{uW}+c_{w} C^{3j*}_{uB{\phi}}]\frac{{\nu} m_{t}}{{\Lambda}^2}
\end{eqnarray}
\begin{eqnarray}
\lambda^{R}_{qt}=\frac{\sqrt{2}}{e}[s_{w}C^{j3}_{uW}+c_{w} C^{j3}_{uB{\phi}}]\frac{{\nu} m_{t}}{{\Lambda}^2}
\end{eqnarray}

For simplification, it has been assumed that the FCNC interaction vertices are non-chiral, meaning ${\lambda}_{qt}^{R}={\lambda}_{qt}^{L}={\lambda}_{q}$
Moreover, it is assumed that ${\lambda}_{ut}$ and ${\lambda}_{uc}$  are equal, as referenced in prior studies. The decay width for the top quark transitioning into $q \gamma)$ via the $tq\gamma$ coupling can be calculated using the following expression: ${\Gamma(t{\rightarrow}q\gamma)}=\frac{\alpha}{2}{\lambda_{q}^2 } {m_{t}}= {0.6763{\lambda_{q}^2 }}$ where $\alpha$ is the fine structure constant, defined as $\alpha=(g_{e}e)^2/(4\pi)=\frac{1}{128.1}$ and $m_t=173.0$ GeV . The branching ratio for this anomalous interaction is then given by:
  \begin{eqnarray}
  BR(t\rightarrow q\gamma)=\frac{\Gamma(t\rightarrow q \gamma)}{\Gamma(t\rightarrow Total)}
  \end{eqnarray}
In the Standard Model, the dominant decay mode of the top quark is $t\rightarrow W b$, , with a decay width of approximately 1.49 GeV, which is used as the total decay width. Both $\Gamma(t\rightarrow W b)$ and $\Gamma(t\rightarrow q \gamma)$ are computed using MadGraph. Thus, the branching ratio for the anomalous coupling can be expressed as the ratio of the $t\rightarrow q\gamma$decay width,(with $\textit{q=u,c}$ ) to the $t\rightarrow W b$ decay width
  \begin{eqnarray}
  BR(t\rightarrow q\gamma)=\frac{\Gamma(t\rightarrow q \gamma)}{\Gamma(t\rightarrow W b)}={0.4573{\lambda_{q}^2}}
  \end{eqnarray} 
 \subsection{Cross Sections}\label{sec3}
 
In this study, we analyzed the process ${\mu} {p} {\rightarrow} {\mu} {\gamma} {p}  {\rightarrow} {\mu} t {\gamma}{\rightarrow} {\mu} W b {\gamma}{\rightarrow} {\mu}{} j j b {\gamma} $  within the framework of the SMEFT. The tree-level Feynman diagrams for the subprocess $ \gamma q (q=u,c) {\rightarrow} t {\gamma}$, which includes contributions from new physics, are displayed in Fig. 1. In both the figures and tables, the process ${\mu} {p} {\rightarrow} {\mu} {\gamma} {p}  {\rightarrow} {\mu} t {\gamma}{\rightarrow}$$ {\mu} W b {\gamma}{\rightarrow} {\mu}{} j j b {\gamma} $ is referred to as the hadronic channel. To simulate the signal and background events, we utilized the Monte Carlo simulation software MadGraph5$\_$aMC@NLO \cite{31}. The FeynRules package \cite{30}, which defines the Standard Model Lagrangian, was extended by incorporating the Universal FeynRules Output (UFO) module \cite{33}, which includes terms for the SMEFT Lagrangian \cite{34}.

\begin{figure}[h]
\centering
\includegraphics[width=0.7\textwidth]{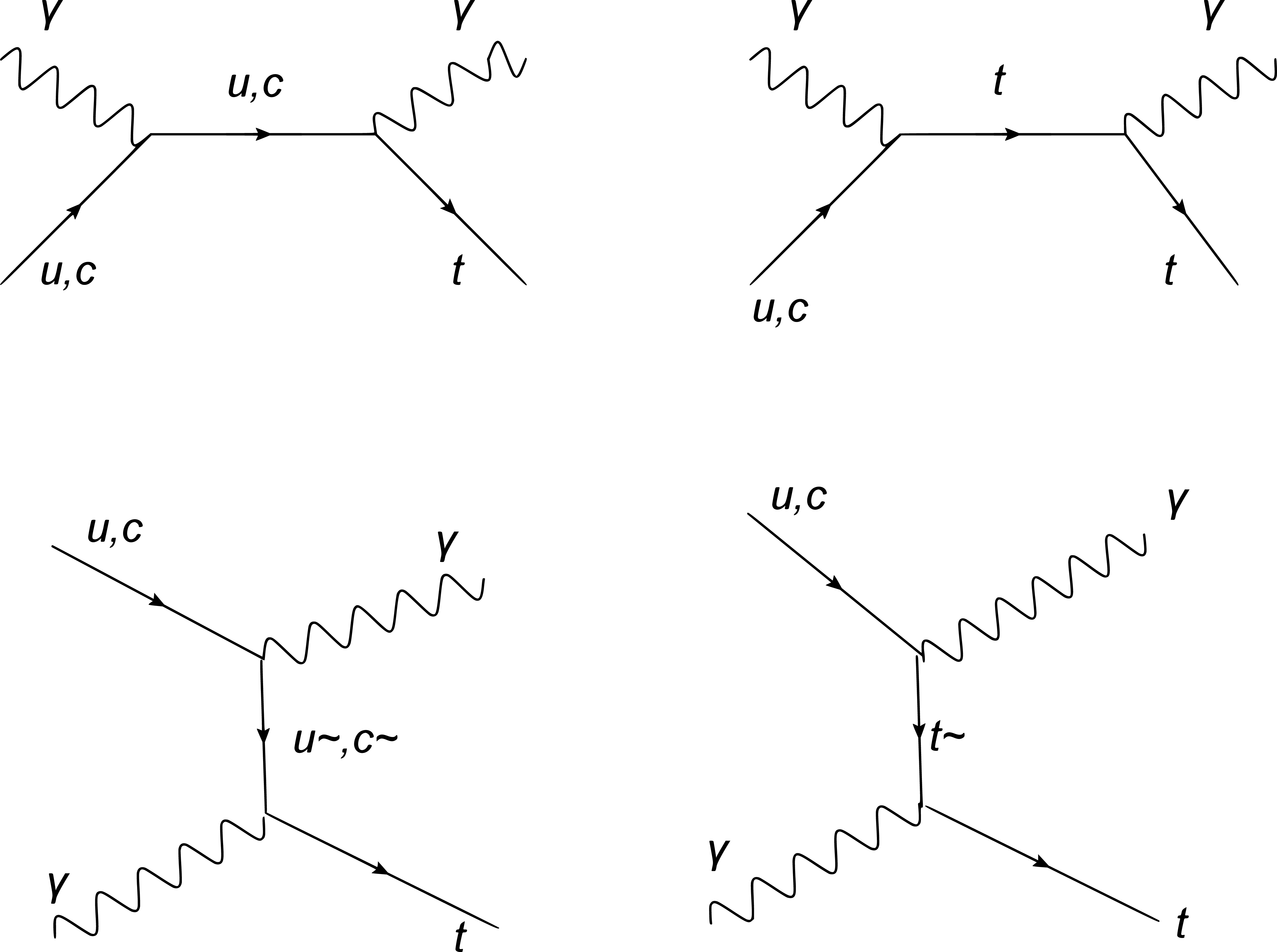}
\caption{ Tree level Feynman diagrams for the subprocesses $ \gamma q {\rightarrow} t  \gamma $ that includes new physics contributions.}
\end{figure}

Here, since the process $ \gamma q {\rightarrow} t \gamma $ is not a process that is within the Standard Model (SM), interactions that result in processes with a final state topology similar to that observed are only considered in the background calculations. Accordingly, these relevant processes are  $\gamma q {\rightarrow} W b \gamma$, $W j \gamma$, $Z j \gamma$. Also, for event analyses in the study, we employed Madanalaysis5 software for both signal and background samples \cite{35}.  Subsequently , Fig1-6 was then plotted considering the basic cut selection. Finally, statistical analyses were performed by applying different cuts to separate the signal and background radiation based on these plots, and the results obtained are listed in Table 1.

\begin{figure}
\includegraphics[width=10.cm]{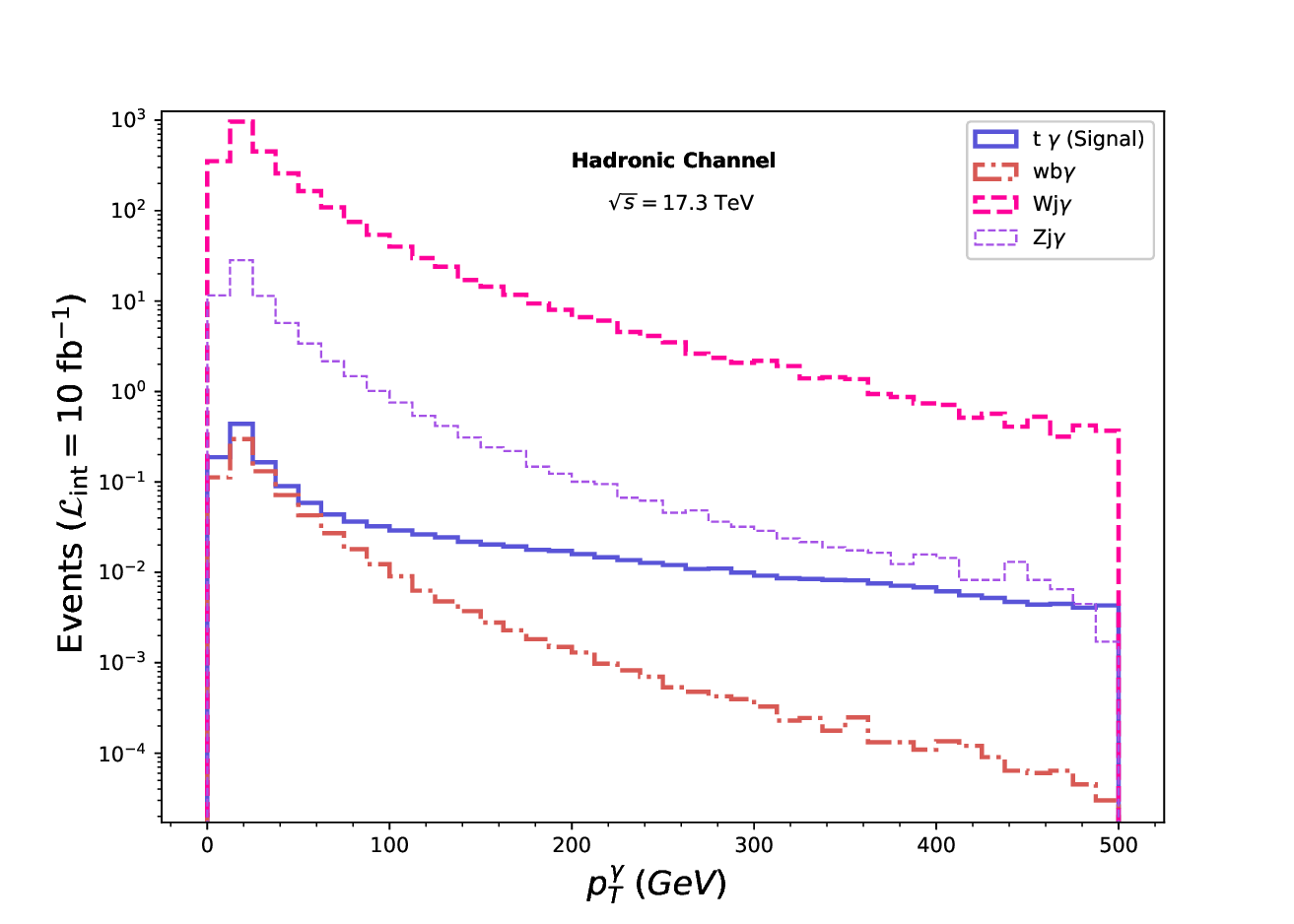}
\caption{$ p_{t}^{\gamma}$ distributions for the signals($t  \gamma $) and SM backgrounds ($W b \gamma$ , $W j \gamma$ ,$ Z j \gamma$)  for center-of mass energy of $\sqrt{s}=17.3TeV$ at the FCC-${\mu}$p.}
\end{figure}
\begin{figure}
\includegraphics[width=10.cm]{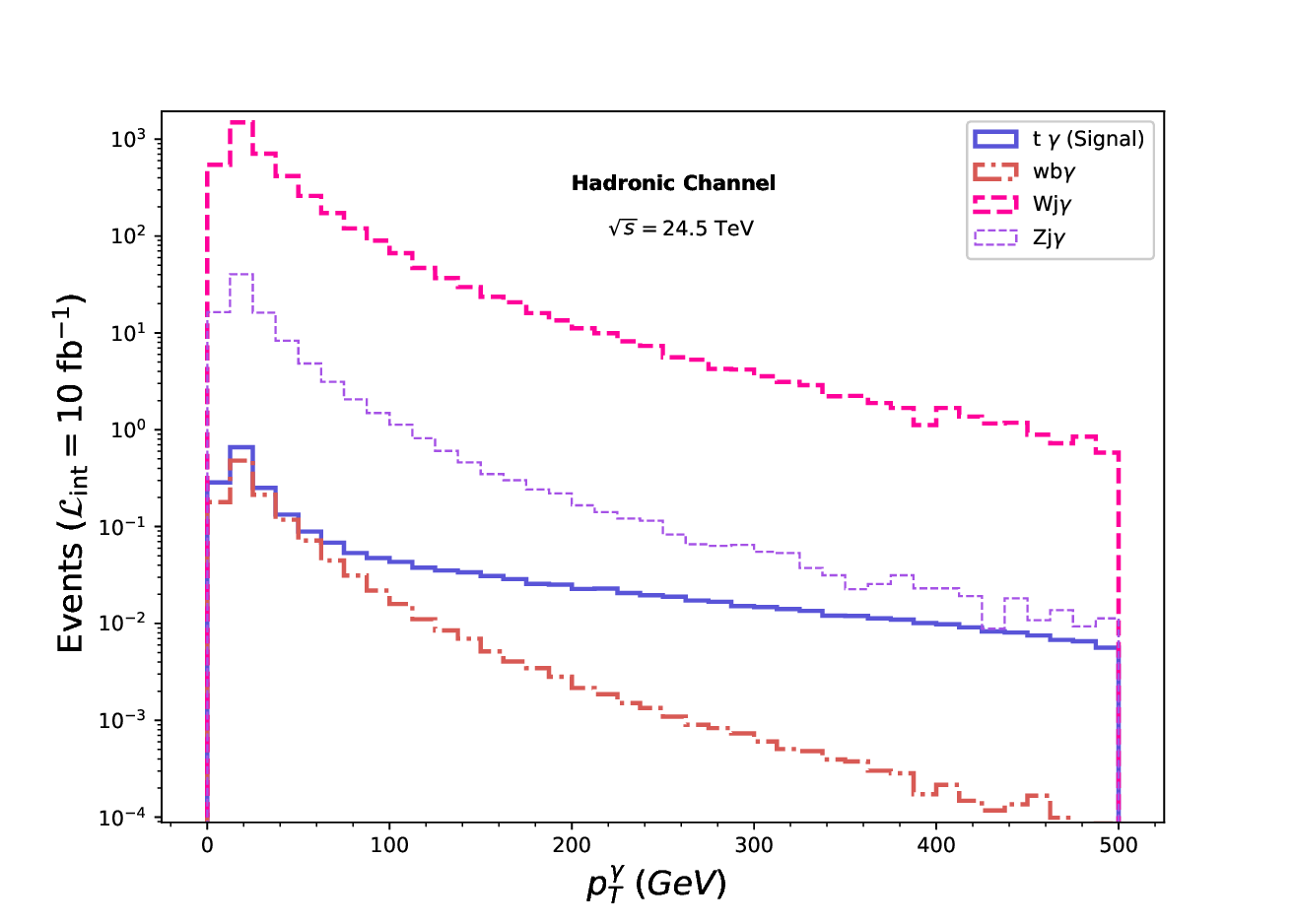}
\caption{Same as Fig3, but for center-of mass energy of $\sqrt{s}=24.5TeV$ at the FCC-${\mu}$p.}
\end{figure}
\begin{figure}
\includegraphics[width=10.cm]{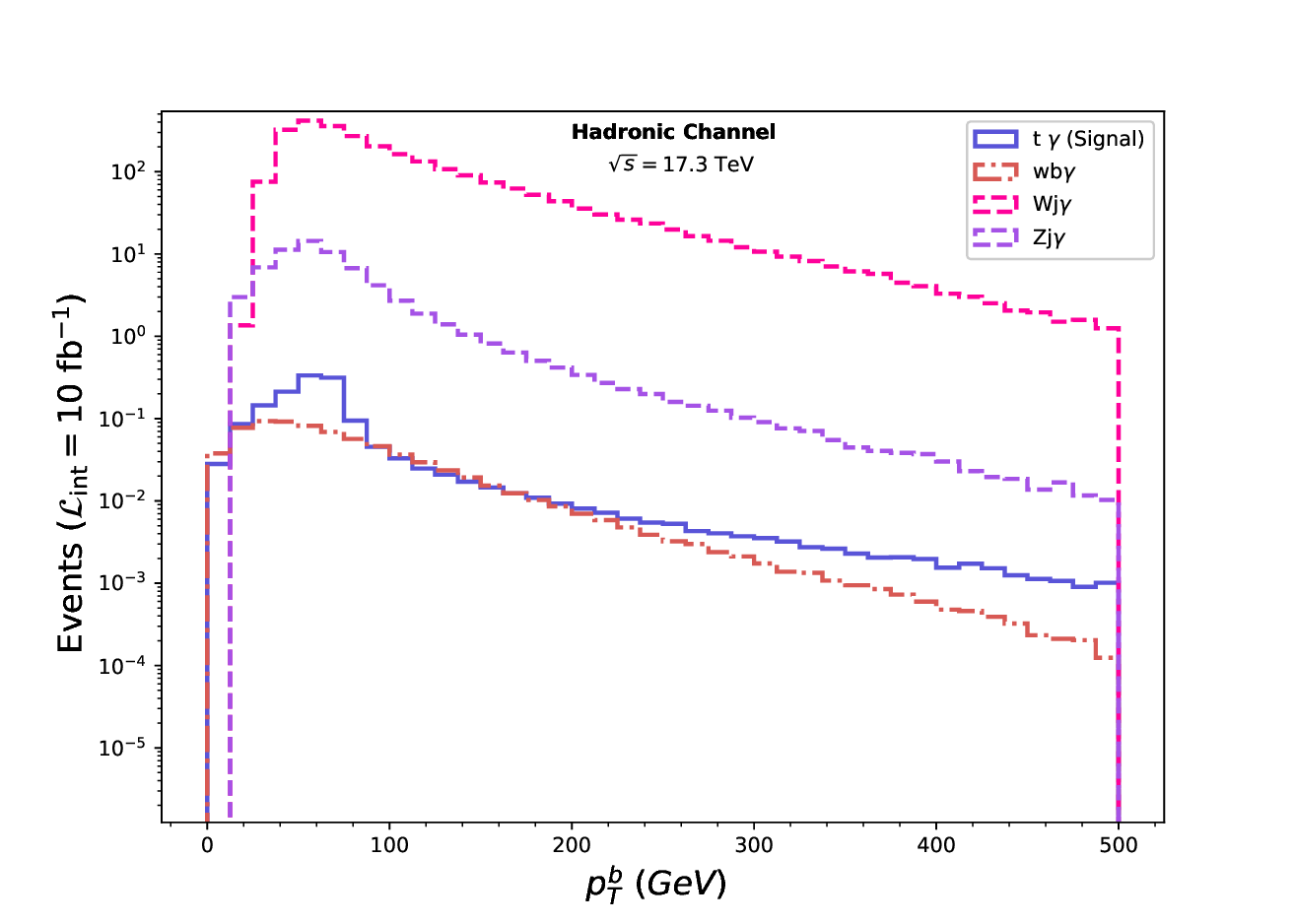}
\caption{$ p_{t}^{b}$ distributions for the signals($t  \gamma $) and SM backgrounds ($W b \gamma$ , $W j \gamma$ ,$ Z j \gamma$)  for center-of mass energy of $\sqrt{s}=17.3TeV$ at the FCC-${\mu}$p.}
\end{figure}
\begin{figure}
\includegraphics[width=10.cm]{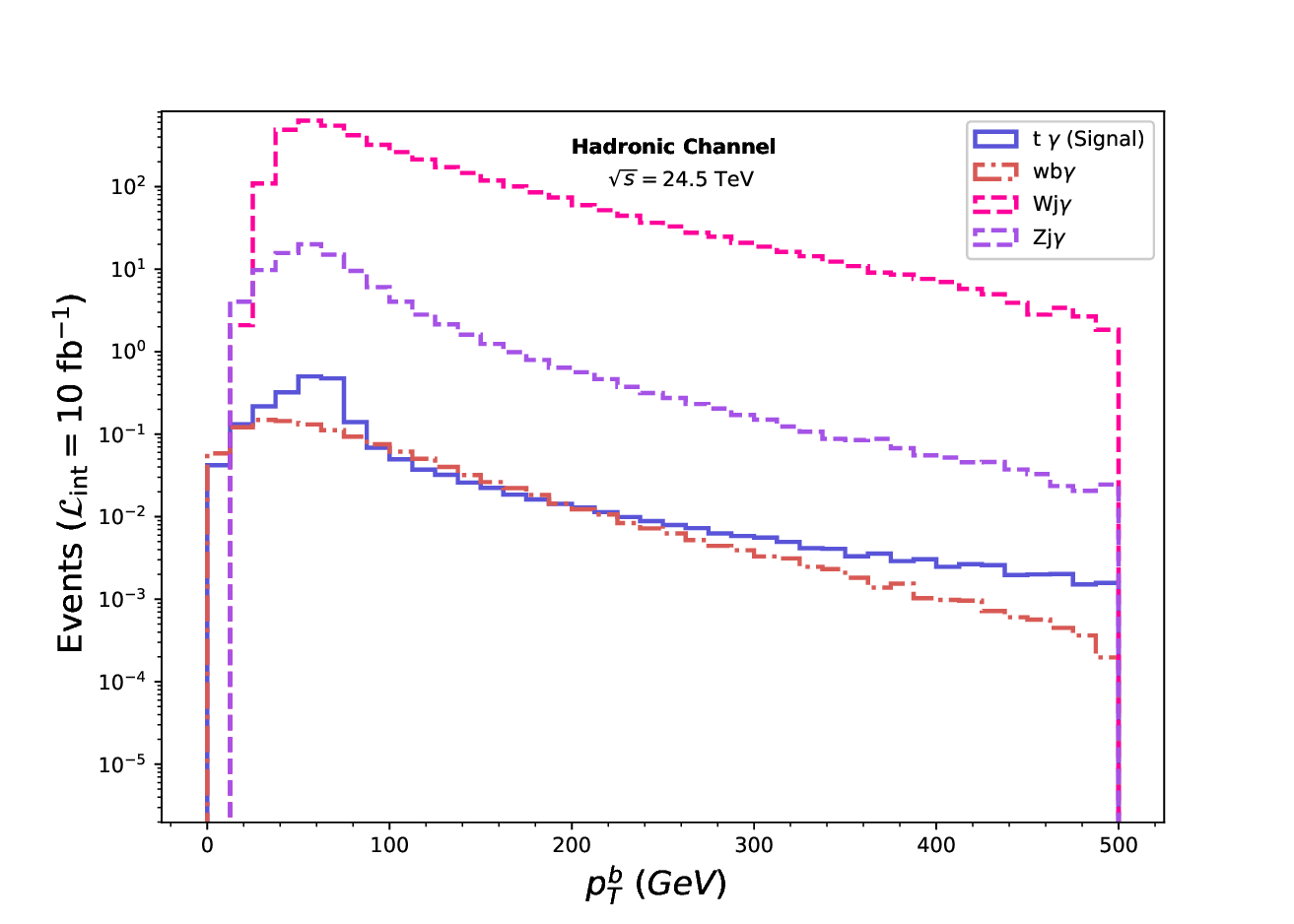}
\caption{Same as Fig5, but for center-of mass energy of $\sqrt{s}=24.5TeV$ at the FCC-${\mu}$p.}
\end{figure}
\begin{figure}
\includegraphics[width=10.cm]{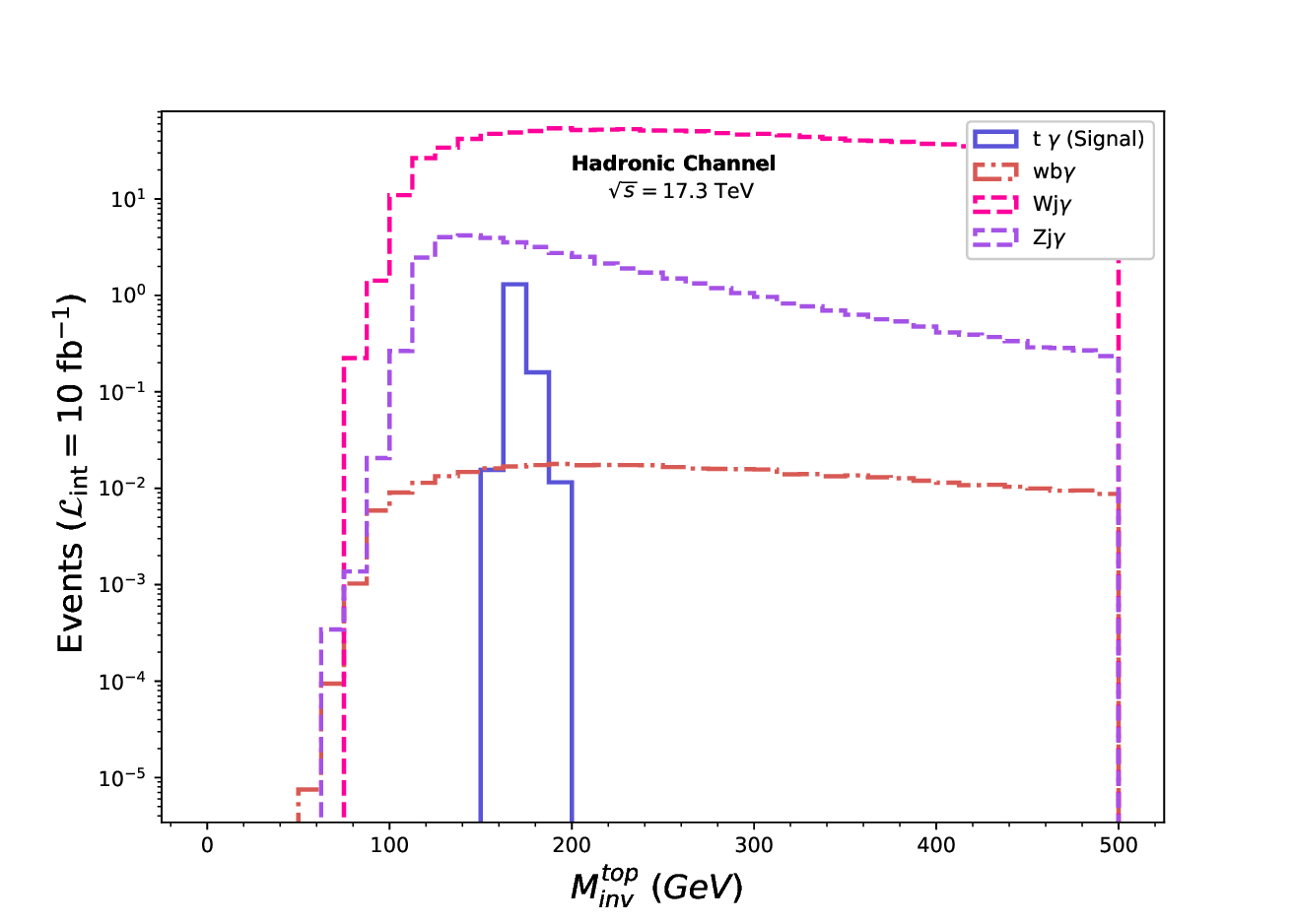}
\caption{ $M_{inv}^{top}$ distributions for the signals($t  \gamma $) and SM backgrounds ($W b \gamma$ , $W j \gamma$ ,$ Z j \gamma$)  for center-of mass energy of $\sqrt{s}=17.3TeV$ at the FCC-${\mu}$p.}
\end{figure}
\begin{figure}
\includegraphics[width=10.cm]{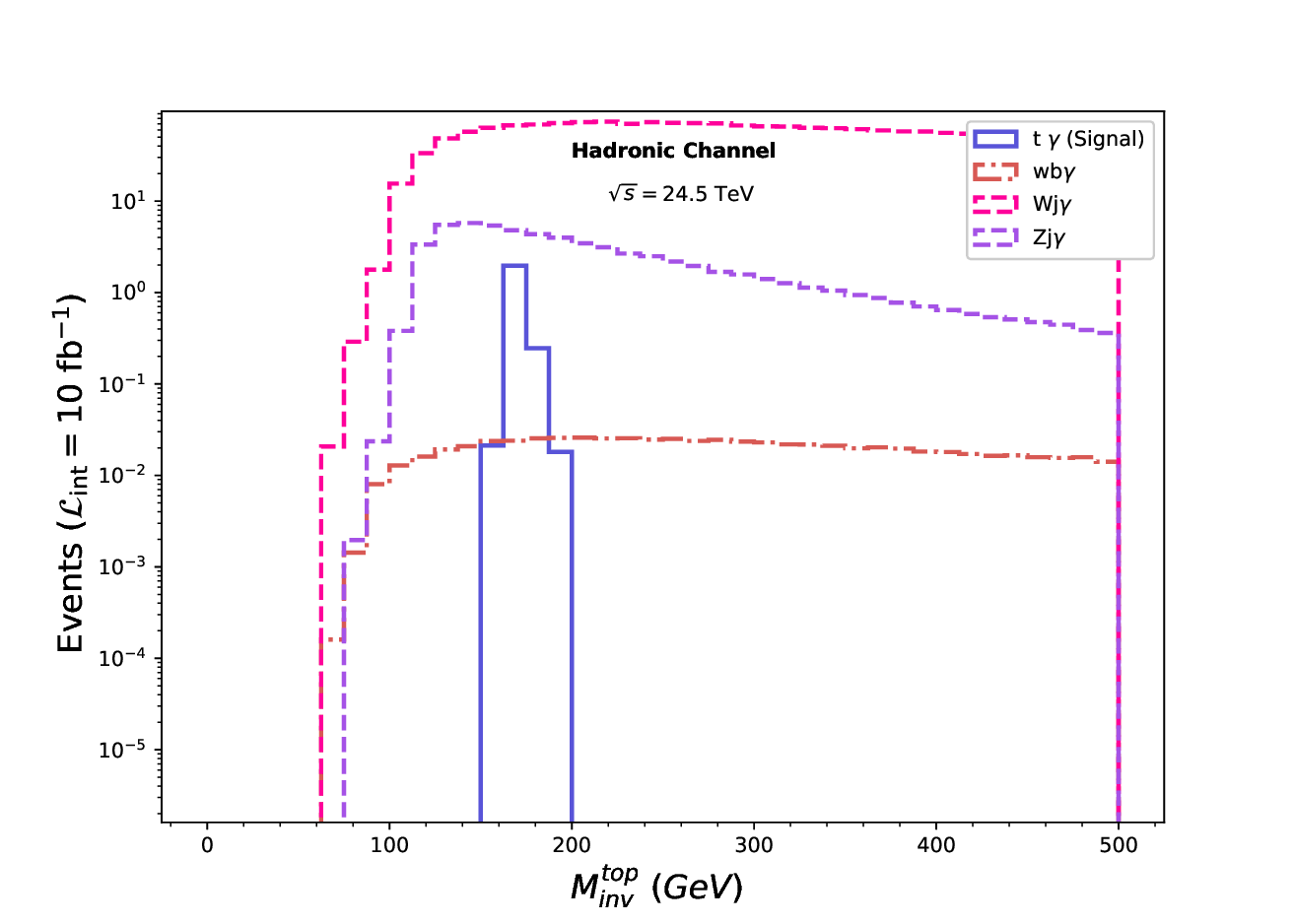}
\caption{Same as Fig6, but for center-of mass energy of $\sqrt{s}=24.5TeV$ at the FCC-${\mu}$p.}
\end{figure}

First, the fundamental cuts were applied using the default settings of madgraph. Then the constraint psedurapidty $\left|{\eta}^{j_{1},j_{2},b,{\gamma}}\right|<2.5 $was added in order to be consistent with the experimental conditions. Furthermore, the transverse momentum cutoff for the final state jets and the b quark is chosen to be $p_{t}^{j_{1},j_{2},b}> 30$ GeV. This choice was made to ensure consistency with experimental conditions and to eliminate low-energy background noise, while still capturing the majority of the signal events. Figures 4 and 5 illustrate the $p_T^b$ distributions for different center-of-mass energies, showing that although a higher $p_T^b$ cut (e.g., 150 GeV) could enhance the signal-to-background ratio, the 30 GeV cut is sufficient to suppress a significant portion of the background, while retaining a substantial number of signal events. This cut allows for a balance between signal preservation and background reduction in the early stages of the analysis. In addition, the transverse momentum cutoffs for the final state particle, the photon, are also introduced. Here, these cutoffs have been decided by analysing Figures 2 and 3. Examining at these figures, the most similar behaviour to our signal process is for the $wb\gamma$ final state, but here we observe that at $p_{T}^{\gamma}> 50 GeV$ the signal process and the $wb\gamma$ process slightly start to diverge from each other, and furthermore, at $p_{T}^{\gamma}> 150 GeV$ the signal process becomes more dominant by strongly eliminating the $wb\gamma$ background. Accordingly, cut1 and cut2 were implemented as two distinct transverse momentum cuts for the photon. As we previously stated, it is assumed that the single  top quark in our signalling process is the w boson and b quark, and this w boson decays hadronically.In this context, it would be useful to introduce a cut4 ($50 GeV< M_{inv}^W<100 GeV$), starting from the assumption that the quarks labelled j1 and j2 are formed by the hadronic decay of the w boson. Accordingly, the distrubition of the number of events with $M_{inv}^{top}$ is presented in the Figures 6 and 7 in order to investigate the behaviour of the background processes and signal process arising from the single produced top quark. As it can be seen in the Figures 6 and 7, the signalling process remains in the range $150 GeV< M_{inv}^{top}<200 GeV$. Choosing this region as the cutoff region is not lead a significant change in the number of events of our signal process, but will significantly reduce the number of background events. Hence, it is consired that it may be appropriate as a final cut to adopt cut5($150 GeV< M_{inv}^{top}<200 GeV$) .

\begin{table}[h]
\begin{center}
\caption{List of applied kinematic cuts for the analysis.}
\begin{tabular}{cccc|cccc}
\hline\hline
Basic cut& $ p_{T}^{\gamma}> 10 GeV$, $ p_{t}^{j}> 20 GeV$ &\\
Cut1&  $p_{T}^{\gamma}> 50$ GeV, $p_{T}^{j_{1},j_{2},b}> 30$ GeV,$\left|{\eta}^{j_{1},j_{2},b,{\gamma}}\right|<2.5 $, $\Delta R=0.7$ & \\
Cut2& Same as cut 1 but $p_{T}^{\gamma}> 150$ GeV    & \\
Cut3&  cut2 + $50 GeV< M_{inv}^W<100 GeV$ &  \\
Cut4&  cut3 + $150 GeV< M_{inv}^{top}<200 GeV$ & \\
\hline\hline
\end{tabular}
\end{center}
\end{table}

The cross sections of the processes $ \gamma q {\rightarrow} t \gamma $, $ wb\gamma $, $ wj\gamma $ and $ Zj\gamma $ are calculated by accepting the cross sections in Table 1. The results of the cross sections calculated at two different centre of mass energy values 17.3 and 24.5 GeV are listed in Tables 2 and 3, respectively. 
\begin{table}
\caption{The cut-based analysis of the cross sections (in pb) for signal and SM background events  at the FCC-${\mu}$p with $\sqrt{s}=17.3 TeV$  assuming  $\lambda_{q}=0.01$.}
\begin{center}
\begin{tabular}{c||c||c||c||cc}
\hline\hline
\multicolumn{6}{c}{$\sqrt{s}=17.3 TeV$} \\
\hline\hline
\cline{1-5}
\multicolumn{1}{c||}{}&\multicolumn{1}{c||}{ signal} &\multicolumn{3}{l}{Background} \\
\hline
cuts&$\gamma q  \rightarrow t \gamma$  &$ wb\gamma$ &$ wj\gamma $  &$ Zj\gamma $& \\
\hline\hline
Basic cut  &$1.491\times10^{-4}$   &$7.531\times10^{-5}$    &$2.635\times10^{-1}$&$6.866\times10^{-3}$&   \\
\hline
kinamatic I  &$3.334\times10^{-5}$   &$3.523\times10^{-6}$    &$1.139\times10^{-2}$&$3.885\times10^{-4}$&   \\
kinamatic II &$2.192\times10^{-5}$   &$5.911\times10^{-7}$    &$1.744\times10^{-3}$&$7.001\times10^{-5}$&   \\
\hline
w recostruction  &$2.183\times10^{-5}$   &$5.86\times10^{-7}$    &$1.362\times10^{-3}$&$3.20\times10^{-5}$&   \\
top recostruction   &$2.183\times10^{-5}$   &$2.28\times10^{-8}$    &$4.13\times10^{-5}$&$5.41\times10^{-6}$&   \\
\hline
\end{tabular}
\end{center}
\end{table}

\begin{table}
\caption{The cut-based analysis of the cross sections (in pb) for signal and SM background events  at the FCC-${\mu}$p with $\sqrt{s}=24.5 TeV$  assuming  $\lambda_{q}=0.01$.}
\begin{center}
\begin{tabular}{c||c||c||c||cc}
\hline\hline
\multicolumn{6}{c}{$\sqrt{s}=24.5 TeV$} \\
\hline\hline
\cline{1-5}
\multicolumn{1}{c||}{}&\multicolumn{1}{c||}{ signal} &\multicolumn{3}{l}{Background} \\
\hline
cuts&$\gamma q  \rightarrow t \gamma$  &$ wb\gamma$ &$ wj\gamma $  &$ Zj\gamma $& \\
\hline\hline
Basic cut  &$2.256\times10^{-4}$   &$1.233\times10^{-4}$    &$4.144\times10^{-1}$&$9.815\times10^{-3}$&   \\
\hline
cut1  &$4.839\times10^{-5}$   &$8.785\times10^{-6}$    &$5.728\times10^{-2}$&$5.314\times10^{-4}$&   \\
cut2 &$3.352\times10^{-5}$   &$1.414\times10^{-6}$    &$1.098\times10^{-2}$&$1.087\times10^{-4}$&   \\
\hline
cut3  &$3.33\times10^{-5}$   &$1.86\times10^{-5}$    &$7.23\times10^{-3}$&$5.06\times10^{-5}$&   \\
cut4   &$3.32\times10^{-5}$   &$3.78\times10^{-8}$    &$9.21\times10^{-5}$&$7.81\times10^{-6}$&   \\
\hline
\end{tabular}
\end{center}
\end{table}

From the tables it can be seen that as a result of all the applied cuts, the cross section of the signal process is reduced by approximately $10^1$, while for the final state processes, this reduction rate is $3*10^3$ for $wb\gamma$, $4*10^3$ for $wj\gamma$ and $1*10^3$ for $Zj\gamma$. This shows that the applied cuts do not cause a significant change in the cross section of our signalling process compared to other processes, but the background processes are strongly suppressed. From this point of view, we analyse ${\mu} {p} {\rightarrow} {\mu} {\gamma} {p} {\rightarrow} {\mu} t {\gamma}{\rightarrow} {\mu} W b {\gamma}{\rightarrow} {\mu}{} \textit{l}{} {\nu}_\textit{l} b {\gamma}$  of the anomalous new physical parameter.

\subsection{Sensitivity at the FCC-${\mu}$p}\label{sec3}

In the field of particle physics, the concept of signal significance is of paramount importance in distinguishing true signals from background noise in experimental data. Signal significance is used to quantify the probability that a detected signal is not just a consequence of random fluctuations. This provides a statistical basis for discovery claims, which is particularly important in high-energy physics investigations, where the goal is often to identify rare events that may be indicative of new physics beyond the Standard Model (BSM). In this context, we apply statistical analysis methods to detect rare events and find evidence for new physics by controlling and accounting for background noise.

To estimate the expected and exclusion significance, denoted as $Z_{excl}$ and $Z_{disc}$, respectively, for small event numbers, the following expressions are used \cite{36}:
\begin{eqnarray}
Z_{disc}=\sqrt{2[(s+b)\ln(\frac{(s+b)(1+{\delta}^{2}b)}{b+{\delta}^{2}b(s+b)})-\frac{1}{{\delta}^2}\ln(1+{\delta}^2\frac{s}{1+{\delta}^{2}b})]}
\end{eqnarray}
\begin{eqnarray}
Z_{excl}=\sqrt{2[s-b\ln(\frac{s+b+x}{2b})-\frac{1}{{\delta}^2}\ln(\frac{b-s+x}{2b})]-(b+s-x)(1+\frac{1}{{\delta}^{2}b})}
\end{eqnarray}

where $ x=\sqrt{(s+b)^{2}-(\frac{4{\delta}^{2}sb^{2}}{1+{\delta}^{2}b}})$. In these equations, s and b represent the number of signal and background events, respectively, while  $ \delta$ refers to the percentage of systematic error on the SM background estimate. For cases where $ \delta \rightarrow 0$ , these expressions can be simplified to:
\begin{eqnarray}
Z_{disc}=\sqrt{2[(s+b)\ln(1+\frac{s}{b})-s]}
\end{eqnarray}

\begin{eqnarray}
Z_{excl}=\sqrt{2[s-b\ln(1+\frac{s}{b})]}
\end{eqnarray}

Regions with $Z_{excl} {\leq} 1.645$ are defined as those that can be excluded at $95\% CL$, regions with $Z_{disc} {\geq} 5$ are considered discovery regions ($5\sigma$), and regions with $Z_{disc} {\geq}2 $ are regarded as weak evidence regions ($2\sigma$).In this paper, we assumed three different cases: $ \delta = 0\%, 5\%, 10\%$.

\begin{table}
\caption{ Bounds on the  $BR(t \rightarrow q\gamma)$ for center-of-mass energy $\sqrt{s}=17.3$ TeV through the processes ${\mu} {p} {\rightarrow} {\mu} {\gamma} {p}  {\rightarrow} {\mu} t {\gamma}{\rightarrow} {\mu} W b {\gamma}{\rightarrow} {\mu}{} j j b {\gamma} $.It is taken under consideration, systematic error as $0\%$,$5\%$,$10\%$ and integrated luminosities as 500,1000,2000,3000,4000,5000 $fb^{-1}$.}
\begin{center}
\begin{tabular}{cc|cccc}
\hline\hline
\multicolumn{6}{c}{$\sqrt{s}=17.3 TeV$} \\
\hline\hline
\multicolumn{2}{c|}{} & \multicolumn{2}{c}{$BR(t \rightarrow q\gamma)$} & \\
\cline{1-6}
${\cal L} \, (fb^{-1})$  & \hspace{0.5cm} $ \delta_{sys}$ \hspace{0.5cm}  &
\hspace{1.5cm} $Z_{disc}(5\sigma)$ \hspace{1.5cm} &\hspace{1.5cm} $Z_{disc}(2\sigma)$ \hspace{1.5cm}&
\hspace{1.5cm} $Z_{excl}$ \hspace{1.5cm} & \\
\hline\hline
500  &  $0\%$   &$1.17\times10^{-4}$    &  $4.29\times10^{-5}$  &$3.70\times10^{-5}$&   \\
500  &  $5\%$   &$1.21\times10^{-4}$  &$4.44\times10^{-5}$   &$3.78\times10^{-5}$&   \\
500  &  $10\%$   &$1.37\times10^{-4}$  & $4.87\times10^{-5}$   &$3.99\times10^{-5}$&   \\
\hline\hline
1000  &  $0\%$   &$7.92\times10^{-5}$    &  $2.98\times10^{-5}$ &$2.53\times10^{-5}$&   \\
1000  &  $5\%$   &$8.55\times10^{-5}$  &  $3.17\times10^{-5}$ &$2.65\times10^{-5}$&   \\
1000  &  $10\%$   &$1.03\times10^{-4}$  & $3.71\times10^{-5}$  &$2.95\times10^{-5}$&   \\
\hline\hline
2000  &  $0\%$   &$5.44\times10^{-5}$    &$2.08\times10^{-5}$  &$1.75\times10^{-5}$&   \\
2000  &  $5\%$   &$6.21\times10^{-5}$  &$2.33\times10^{-5}$  &$1.92\times10^{-5}$&   \\
2000  &  $10\%$   &$8.26\times10^{-5}$  &$2.99\times10^{-5}$  &$2.31\times10^{-5}$&   \\
\hline\hline
3000  &  $0\%$   &$4.38\times10^{-5}$    & $1.68\times10^{-5}$  &$1.42\times10^{-5}$&   \\
3000  &  $5\%$   &$5.25\times10^{-5}$  & $1.98\times10^{-5}$  &$1.62\times10^{-5}$&   \\
3000  &  $10\%$   &$7.49\times10^{-5}$  &  $2.72\times10^{-5}$ &$2.08\times10^{-5}$&   \\
\hline\hline
4000  &  $0\%$   &$3.76\times10^{-5}$    &$1.46\times10^{-5}$ &$1.22\times10^{-5}$&   \\
4000  &  $5\%$   &$4.72\times10^{-5}$  &$1.79\times10^{-5}$  &$1.45\times10^{-5}$&   \\
4000  &  $10\%$   &$7.08\times10^{-5}$  &$2.57\times10^{-5}$  &$1.95\times10^{-5}$&   \\
\hline\hline
5000  &  $0\%$   &$3.34\times10^{-5}$    &$1.30\times10^{-5}$ &$1.08\times10^{-5}$&   \\
5000  &  $5\%$   &$4.38\times10^{-5}$  &$1.66\times10^{-5}$&$1.34\times10^{-5}$&   \\
5000  &  $10\%$   &$6.82\times10^{-5}$  &$2.48\times10^{-5}$    &$1.86\times10^{-5}$&   \\
\hline\hline
\end{tabular}
\end{center}
\end{table}

\begin{table}
\caption{Bounds on the  $BR(t \rightarrow q\gamma)$ for center-of-mass energy $\sqrt{s}=24.5$ TeV through the processes ${\mu} {p} {\rightarrow} {\mu} {\gamma} {p}  {\rightarrow} {\mu} t {\gamma}{\rightarrow} {\mu} W b {\gamma}{\rightarrow} {\mu}{} j j b {\gamma} $.It is taken under consideration, systematic error as $0\%$,$5\%$,$10\%$ and integrated luminosities as 500,1000,2000,3000,4000,5000 $fb^{-1}$.}
\begin{center}
\begin{tabular}{cc|cccc}
\hline\hline
\multicolumn{6}{c}{$\sqrt{s}=24.5 TeV$} \\
\hline\hline
\multicolumn{2}{c|}{} & \multicolumn{2}{c}{$BR(t \rightarrow q\gamma)$} & \\
\cline{1-6}
${\cal L} \, (fb^{-1})$  & \hspace{0.5cm} $ \delta_{sys}$ \hspace{0.5cm}  &
\hspace{1.5cm} $Z_{disc}(5\sigma)$ \hspace{1.5cm} &\hspace{1.5cm} $Z_{disc}(2\sigma)$ \hspace{1.5cm}&
\hspace{1.5cm} $Z_{excl}$ \hspace{1.5cm} & \\
\hline\hline
500  &  $0\%$   &$1.06\times10^{-4}$    &$2.81\times10^{-5}$&$3.41\times10^{-5}$&   \\
500  &  $5\%$   &$1.15\times10^{-4}$  &$2.91\times10^{-5}$&$3.58\times10^{-5}$&   \\
500  &  $10\%$   &$1.37\times10^{-4}$  &$3.19\times10^{-5}$&$4.01\times10^{-5}$&   \\
\hline\hline
1000  &  $0\%$   &$7.32\times10^{-5}$    &$1.95\times10^{-5}$&$2.36\times10^{-5}$&   \\
1000  &  $5\%$   &$8.41\times10^{-5}$  &$2.07\times10^{-5}$&$2.60\times10^{-5}$&   \\
1000  &  $10\%$   &$1.13\times10^{-4}$  &$2.43\times10^{-5}$&$2.77\times10^{-5}$&   \\
\hline\hline
2000  &  $0\%$   &$5.01\times10^{-5}$    &$1.36\times10^{-5}$&$1.64\times10^{-5}$&   \\
2000  &  $5\%$   &$6.43\times10^{-5}$  &$1.53\times10^{-5}$&$1.97\times10^{-5}$&   \\
2000  &  $10\%$   &$9.73\times10^{-5}$  &$1.96\times10^{-5}$&$2.68\times10^{-5}$&   \\
\hline\hline
3000  &  $0\%$   &$4.10\times10^{-5}$    &$1.10\times10^{-5}$&$1.33\times10^{-5}$&   \\
3000  &  $5\%$   &$5.64\times10^{-5}$  &$1.30\times10^{-5}$&$1.72\times10^{-5}$&   \\
3000  &  $10\%$   &$9.16\times10^{-5}$  &$1.78\times10^{-5}$&$2.50\times10^{-5}$&   \\
\hline\hline
4000  &  $0\%$   &$3.52\times10^{-5}$    &$9.53\times10^{-6}$&$1.15\times10^{-5}$&   \\
4000  &  $5\%$   &$5.21\times10^{-5}$  &$1.17\times10^{-5}$&$1.58\times10^{-5}$&   \\
4000  &  $10\%$   &$8.86\times10^{-5}$  &$1.68\times10^{-5}$&$2.40\times10^{-5}$&   \\
\hline\hline
5000  &  $0\%$   &$3.14\times10^{-5}$    &$8.50\times10^{-6}$ &$1.02\times10^{-5}$&   \\
5000  &  $5\%$   &$4.93\times10^{-5}$  &$1.09\times10^{-5}$&$1.49\times10^{-5}$&   \\
5000  &  $10\%$   &$8.67\times10^{-5}$  &  $1.66\times10^{-5}$ &$2.35\times10^{-5}$&   \\
\hline\hline
\end{tabular}
\end{center}
\end{table}

Tables 4 and 5 present the sensitivity analysis for two different centre-of-mass energies ($\sqrt{s}=17.3 TeV$ and $\sqrt{s}=24.5 TeV$) to determine the $BR(t \rightarrow q\gamma)$ limits on the branching ratio of this experimentally rarely observed process. The factors considered here are the systematic error rate ($ \delta_{sys}$) and different values of the integrated luminosity ${\cal L}$. By performing different limit searches in the tables, both discovery limits and exclusion limits can be obtained. Here, the limit $Z_{exp}(5\sigma)$ denotes the branching ratio at which a discovery can be made for the $t \rightarrow q\gamma$ decay. Among the limits, the largest values are seen here, because a higher number of events is required to make a discovery. For example, in Table 4, under the assumption that $ \delta_{sys}=0\%$, the discovery limit for $500fb^{-1}$ is $1.17\times10^{-4}$, while for 5000 $fb^{-1}$ it is $3.34\times10^{-5}$. It can be seen that even at elevated levels of irradiation, the branching ratio of the established limit for the discovery remains quite small and this situation indicates that the decay is a rare occurrence. The limit $Z_{exp}(2\sigma)$ is the minimum branching ratio limit at which the event $t \rightarrow q\gamma$ can be detected and  is at the two sigma level in terms of statistical significance. It provides more stringent bounds than $(5\sigma)$. For instance, for 500 $fb^{-1}$ the $(2\sigma)$ limit is $4.29\times10^{-5}$, while for 5000 $fb^{-1}$ the limit is $1.30\times10^{-5}$. Accordingly, although the $(2\sigma)$ limit does not theoretically imply discovery, it is a critical limit for a potential detection. 

The $Z_{excl}$ exclusion limit is used to exclude the branching ratio of  $t \rightarrow q\gamma$ when no signal is observed in the experiment. Comparing Tables 4 and 5, the most stringent exclusion limit is obtained as $1.02\times10^{-5}$ for 5000 $fb^{-1}$ at the centre-of-mass energy $\sqrt{s}=24.5 TeV$. The absence of a signal in experimental studies requires the theoretical models to have BR values below these limits. Accordingly, a comparison of our results with the latest experimental data obtained by the CMS group shows that the most stringent limits obtained in this study are 2.5 times more restrictive than the experimental limits.

Tables 4 and 5 are presented for two different centre-of-mass energies $\sqrt{s}=17.3 TeV$ and $\sqrt{s}=24.5 TeV$, respectively. A comparison of the two tables reveals that the limits obtained for the high value of the centre-of-mass energy, 24.5 TeV, are all greater than those obtained for the lower value, 17.3 TeV as expected. This behaviour can be attributed to the fact that an increase in the centre-of-mass energy increases the observability of rare decay processes of high-mass particles such as top quarks. This allows testing of rare processes that were previously impossible to observe.
The tables also present the probability limits for the $t \rightarrow q\gamma$ decay under different conditions of integrated luminosity ${\cal L}$ and systematic error ($ \delta_{sys}$). This allows us to understand the effect of experimental sensitivity and uncertainties on the branching ratio limits. In this context, we first examine the effect of luminosity values on the branching ratio limits. As it is known, luminosity represents the amount of data collected at a given collision energy, and the high luminosity leads to more events observed and reduced statistical uncertainty. Accordingly, it is evident that the branching ratio limits tighten with increasing irradiance. Table 4, which pertains to a collision energy of 24.5 TeV, shows that the discovery limit is $3.41\times10^{-5}$ for the lowest systematic error acceptance, 500 $fb^{-1}$, whereas at 5000 $fb^{-1}$, this limit is reduced to $1.02\times10^{-5}$. This exemplifies that the limits improve by approximately threefold when the number of events is increased by a factor of ten. This ratio indicates how experimental precision improves directly with the amount of data. Thus, at high luminosity, both the discovery and exclusion limits are expected to become more precise.
In the context of the systematic error effect, it can be said that $ \delta_{sys}$ directly affects the branching ratio limits as a consequence of uncertainties in the experimental measurements. Here, when $ \delta_{sys}=0\%$, only statistical uncertainty is taken into account, while systematic error rates of ($ \delta_{sys}=5\%$) and ($ \delta_{sys}=10\%$) represent possible additional uncertainties in the experiment. For example, in the $(2\sigma)$ region at the maximum luminosity value of 5000 $fb^{-1}$, the branching ratio limits obtained from statistical calculations are $8.50\times10^{-6}$ for $ \delta_{sys}=0\%$ and $1.66\times10^{-5}$ for $ \delta_{sys}=10\%$. This suggests that systematic errors directly expand the branching ratio limits, while experimental uncertainties have a relaxing effect on the limits. Accordingly, it can be inferred that minimising systematic uncertainties will play a key role in obtaining tighter bounds. Hence, it can be concluded that not only increasing the luminosity is not enough for successful experiments, but also systematic errors should be managed effectively.
In summary, increasing the energy of the centre of mass makes an important contribution to the search for new physics by increasing the sensitivity. Also, increasing luminosity stringent the limits of rare processes. However, for experiments to be successful, systematic errors must be managed along with high luminosity. In this regard, reducing systematic errors is vital to enhance the reliability of experiments. The findings in the relevant two tables provide a comprehensive framework for analysing the sensitivity of experiments under different energy and error conditions, offering valuable insights into experimental and theoretical implications.

\section{Conclusion}
In this study, we analyzed anomalous Flavor-Changing Neutral Current (FCNC) interactions related to the top quark, specifically the $tq\gamma$ transition, using the SMEFT framework at the FCC-${\mu}$p. These processes are highly suppressed in the SM and serve as a clean probe for new physics beyond the SM. Our results indicate that the FCC-${\mu}$p collider, with its high energy and luminosity capabilities, significantly improves sensitivity to these rare processes, offering discovery potential far exceeding current collider limits. Specifically, our results show that the branching ratios for these processes can be constrained to the order of $10^{-5}$, which is approximately 30\%. more stringent than the current experimental limits from the ATLAS and CMS experiments. 

Moreover, we demonstrated that higher center-of-mass energies and integrated luminosities at the FCC-${\mu}$p collider enhance the discovery potential, not only tightening exclusion limits but also improving the chances of detecting new physics signals. The analysis of cross sections and statistical significance further confirmed that the FCC-${\mu}$p is capable of probing these rare decays, which would be inaccessible at existing colliders. These findings highlight the essential role of future high-energy colliders in exploring the flavor structure of the top quark and in searching for new physics beyond the Standard Model. This study contributes to the growing body of work that positions FCNC processes involving the top quark as a key focus for future experimental and theoretical investigations.

\end{document}